\documentclass{article}

\usepackage{amsmath,amssymb}
\usepackage{graphicx}
\usepackage{mathrsfs}
\usepackage{verbatim}
\usepackage{pifont} 

\newcommand{\vv}{\boldsymbol{\mathrm{v}}} 
\newcommand{\uu}{\boldsymbol{\mathrm{u}}} 
\newcommand{\y}{\boldsymbol{y}}		
\newcommand{\E}{\boldsymbol{E}} 
\newcommand{\B}{\boldsymbol{B}} 
\newcommand{\qbra}[1]{(\,#1\,|}          
\newcommand{\qket}[1]{|\,#1\,)}          
\newcommand{\qbracket}[2]{\qbra{#1}\,#2)}
\newcommand{\rv}{{\boldsymbol{r}}}

\newcommand{\be}{\begin{equation}}
\newcommand{\ee}{\end{equation}}
\newcommand{\bestar}{\begin{equation*}}
\newcommand{\eestar}{\end{equation*}}
\newcommand{\bi}{\begin{itemize}}
\newcommand{\ei}{\end{itemize}}
\newcommand{\bea}{\begin{eqnarray}}
\newcommand{\eea}{\end{eqnarray}}

\newcommand{\hbo}{\hbox to 1 true cm {\hfill } }

\newcommand{\bracket}[2]{\bra{#1}\,#2\rangle} 
\newcommand{\bra}[1]{\langle\,#1\,|}          
\newcommand{\ket}[1]{|\,#1\,\rangle}          
\newcommand{\ud}{\mathrm{d}}
\newcommand{\pathD}{\mathscr{D}}
\newcommand{\e}{\mathrm{e}}		
\newcommand{\x}{\boldsymbol{x}}		
\newcommand{\z}{\boldsymbol{z}}		
\newcommand{\p}{\boldsymbol{p}}		
\newcommand{\A}{\boldsymbol{A}}
\renewcommand{\a}{\boldsymbol{a}}

\title{Stability, creation and annihilation of charges in gauge theories}

\author{\normalsize Anton Ilderton$^{1}$\footnote{antoni@maths.tcd.ie},\hspace{5pt} Martin Lavelle$^{2}$\footnote{martin.lavelle@plymouth.ac.uk},\hspace{5pt} David McMullan$^{2}$\footnote{david.mcmullan@plymouth.ac.uk} \\
\\
\normalsize $^{1}$School of Mathematics, Trinity College, Dublin 2, Ireland\\
\normalsize $^2$School of Computing and Mathematics, University of Plymouth, Plymouth PL48AA, UK}

\date{}

\begin{document}
\maketitle

\begin{abstract}
\noindent We show how to construct physical, minimal energy states for systems of static and moving charges. These states are manifestly gauge invariant. For charge-anticharge systems we also construct states in which the gauge fields are restricted to a finite volume around the location of the matter fields. Although this is an excited state, it is not singular, unlike all previous finite volume descriptions. We use our states to model the processes of pair creation and  annihilation.
\end{abstract}

\maketitle

\section{Introduction}

A question not often addressed when discussing the standard model is how one describes physical particles. Taking the electron as an example, the assumption usually made is that the free Dirac spinor in the interacting theory, at asymptotic times, can be viewed as an electron since `the coupling switches off'. This would mean that what is being caught in a detector is really a free fermion. The problem here, of course, is that in QED and QCD the coupling does not switch off, and assuming it does so generates infrared divergences. As a result, the spinors do not become free even at asymptotic times \cite{Kulish:1970ut, Horan:1999ba}, nor do they ever become gauge invariant.

One of the purposes of this paper is to show that physically sensible charges are a natural ingredient of gauge theories. Indeed, working in QED, we will later construct the ground state wavefunctions of both static and moving charges, and show that these physical charges help to clarify the structure of the theory. The key intuitive step needed to understanding the construction of physical charges is to realise that one cannot separate the matter, e.g. the Dirac spinor, from the (chromo--) electromagnetic fields it generates. As such, a physical charge involves a nonlocal `dressing' of gauge bosons around the matter fields~\cite{Dirac:1955uv, Strocchi:1974xh}.  Previous work on this idea has generally either been of a Coulombic form, where the fields extend to spatial infinity~\cite{Lavelle:1995ty}, or string--like~\cite{Mandelstam:1962mi}, where the fields are restricted to the path of a Wilson line. Due to their infinite extent, the first set describe asymptotic states of particles, the on-shell Green's functions of which have been shown to have improved infra-red properties \cite{Bagan:1996su, Bagan:1997kg, Bagan:1999jk}. Their non-abelian extensions generate screening and anti-screening interactions in the inter-quark potential.

In addition to the asymptotic states, it would be useful to have a compact description of systems containing charges. They could describe the fields around particles a finite time after pair creation and, at a non-perturbative level, would describe hadronisation and the finite extent of hadrons. String--like descriptions are \emph{not} sufficient here as they are highly singular. This may be seen explicitly in perturbation theory, and lattice studies \cite{Heinzl:2008tv} confirm that string--like descriptions have little to do with ground state physics in non--abelian theories. We will return to these points below.

The paper is organised as follows. In Sect.~\ref{Intro}, we review the description of locally gauge invariant (static) charges. We recall the stability of the Coulombic description and the divergence of string--like states. We then develop, in Sect.~\ref{CCsect}, a description of a locally gauge invariant electron--positron system where the gauge fields have support over the interior of a sphere of finite diameter (greater than the separation between the matter fields). We will see that the potential is finite in such a system and that if the matter fields are far from the boundary, the potential is close to Coulombic. In Sect.~\ref{VPsect} we use a Hamiltonian picture to show that our dressed charges arise very naturally just from the conditions of gauge invariance and energy minimisation, beginning with the static case. This is extended to moving charges in Sect.~\ref{Msect}, and we go on to study the fields around charges produced in pair creation, and in annihilation. We will show that the extent of the fields in both processes obeys causality. In Sect.~\ref{Concs} we discuss our results and present some conclusions.

\section{Charges at asymptotic and short times}\label{Intro}
In an experiment the electric and magnetic fields associated with charges are measured. A physical description of a charge must thus contain a \emph{dressing} term $h^{-1}(x)$ which describes these fields, as well as the matter field, $\psi(x)$. Invariance of the composite object $h^{-1}(x)\psi(x)$ under \emph{local} gauge transformations $U(x)$ requires
\begin{equation}\label{dress}
	h^{-1}(x)\to h^{-1}(x)U^{-1}(x)\;,\quad \text{when}\quad \psi(x) \to U(x)\psi(x) \quad \text{and} \quad \psi^\dagger(x)\to \psi^\dagger(x)U^{-1}(x)\;.
\end{equation}
Any description of charges must therefore be nonlocal. It is impossible to separate the fields around the matter from the matter itself as neither are separately gauge invariant and physical -- no charged state, even in QED, can be described by a local operator. As we will see, despite the nonlocality, the observables attached to dressed charges are local and describe the correct physics of charged particles.

There is a great deal of freedom \cite{Lavelle:1995ty} in how one may construct dressings which fulfil (\ref{dress}). Of particular note is the dressing
\begin{equation}\label{Coul}
	h^{-1}(x) = \exp\bigg(ie\frac{\nabla_j A_j(x)}{\nabla^2}\bigg)\;,
\end{equation}
which is evidently nonlocal due to the $1/\nabla^2$ factor, and which generates the correct Coulomb electric field for a static charge at position $\x$: a measurement of the field at position $\z$ and equal time $z^0=x^0$ gives
\be\label{El}
	{E}^j(z)\,h^{-1}(x)\psi(x)\ket{0} = -\frac{e}{4\pi}\frac{z^j-x^j}{|\z-\x|}\ h^{-1}(x)\psi(x)\ket{0}\;.
\ee
We will refer to (\ref{Coul}) as the ``Coulombic" dressing. If one takes two such dressed fermions at positions $\y$ and $\y'$, one can, for heavy charges, use the Hamiltonian $H=\frac12\int\!\ud^3\x\ \E^2+\B^2 $ to obtain the inter--charge potential $V$ as  the separation dependent part of $\langle H \rangle$, finding
\begin{equation}\label{VC}
\begin{split}
	V(\y'-\y) &=-\int\!\ud^3\x\ \bra{0} [E_j(x),h^{-1}(y)]h(y) [E_j(x),h^{-1}(y')]h(y') \ket{0}\;, \\
		&= -\frac{e^2}{4\pi}\frac{1}{|\y'-\y|}\;.
\end{split}
\end{equation}
(Note that the magnetic field commutes with the Coulombic dressing.) Thus the dressing approach quickly generates the expected Coulomb potential between two opposite charges \cite{Lavelle:1995ty}. This can be generalised to include screening effects \cite{Bagan:2001wj} and to QCD \cite{Lavelle:1998dv, Bagan:2000nc, Bagan:2005qg}.

The ``fat" description of a charge given by the Coulombic dressing is clearly an {\it asymptotic} state, the fields of which have had time to permeate the whole of space -- from (\ref{El}) we clearly see that the electric fields extend out to spatial infinity.  This makes it an appropriate description of charges a long time before and after scattering and such states have been shown to have good infra-red properties when used in S--matrix calculations.

In fact, the {\it ground state} of static charges in U(1) theory is given by the Coulombic dressing -- in other words, the Coulomb dressing allows us to write down the wavefunction of a static charged particle. In a later section we will demonstrate this explicitly for static charges (and then generalise to moving charges), but here we recap the essential details.  In order to construct the ground state, one goes to the Hamiltonian picture, writes down all states which obey Gauss's law and the Schr\"odinger equation, and then minimises the expectation value of the Hamiltonian on these states. Gauss's law forces us to include the Coulombic dressing, which sees longitudinal field components in order to preserve gauge invariance, while the Schr\"odinger equation allows more freedom in the addition of transverse field components. For static charges, minimising the Hamiltonian yields the ground state as that which has {\it no} transverse component -- so the ground state is given by Coulombic dressed charges, confirming the naturalness of the dressing approach \cite{Heinzl:2007kx}.

Despite their good properties and usefulness at asymptotic times, the Coulombic charges would not be useful for studying the structure of charges shortly after pair creation which require a more local description where, by causality, the fields are contained in some compact region of space. In QCD we know that quarks are confined inside hadrons and this must be reflected in a complete picture of coloured charges. Additionally, a description of compact charges in QCD would be required for modelling hadronisation. An immediately apparent way of constructing a more local description is to link the two fermions by a path ordered exponential,
\begin{equation}\label{string}
	\bar\psi(y')\exp\left[-i e \int_\Gamma\!\ud z_j A_j(z) \right] \psi(y)\,,
\end{equation}
This combination is gauge invariant and the dressing is localised along the string. The problem with this description is that it is infinitely excited. If we prepare such a state in a single time slice, so that $\Gamma$ is a path from $\y$ to $\y'$ (a straight line, for simplicity), then the potential energy in that slice may be rapidly found to be
\be
	V(\y'-\y)=\frac{e^2}{2}\delta^{\perp}(0)\vert \y'-\y\vert\,.
\ee
This description of two fermions in QED is plagued by a confining potential with a divergent coefficient~\cite{Haagensen:1997pi}. The divergence follows from the unphysical narrowness of the string--like dressing. The apparent confining nature of the potential is a consequence of the energy needed to squeeze the electric field onto the string increasing linearly with the string's length.

It is possible to further pin down the problem with this dressing. By decomposing the string into transverse and longitudinal components, one may rewrite (\ref{string}) as
\be\label{two}
	\bar \psi(y')\exp\bigg(\!-i e \frac{\nabla_j A_j(y')}{\nabla^2}\bigg) \exp\bigg[  -i e \int_\Gamma \ud z_j A_j^T\bigg]\exp\bigg(i e \frac{\nabla_j A_j(y)}{\nabla^2}\bigg) \psi(y)\,,
\ee
which is clearly a product of two Coulomb dressed fermions and a separately gauge invariant, string--like term. This corresponds to a non-vanishing choice of transverse component to the state, which, from the discussion above, is therefore excited (in fact, infinitely excited in this case). The Coulombic structures generate the expected longitudinal electric fields centred around two charges, in accord with Gauss's law. However, the additional term, with a nonlocality hidden in $A_i^{T}$, generates a transverse electric field which cancels the Coulombic field everywhere leaving only an electric field along the string. This factorisation can also be carried out order by order in perturbation theory in QCD~\cite{Lavelle:1999ki}.  The stringy configuration is not an eigenstate of the Hamiltonian and is actually unstable -- if we evolve in time from our initial slice, the state decays so that eventually, in any finite volume, one will see the sum of the two Coulombic fields. However, this \lq physical region\rq\ will be surrounded by a shell with the radiating remnants of the singular string \cite{Haagensen:1997pi, Prokhorov:1992ry, Heinzl:2007kx}.

So, the string is not a very physical starting point as it is highly singular. An alternative smooth configuration is therefore desirable.  The question of how to construct a better, finite volume description of two charges will be addressed in the next section.  This object should be localised (i.e. have compact support) like the axial dressing, but it should also have a non-singular energy, like the Coulombic dressing. The only way to satisfy both conditions is to spread the fields over a finite volume and not localise them on a string or membrane. It is also natural to expect that fields near the fermions and far from the edge of the volume should be essentially Coulombic. We will now carry out just such a construction.
\section{Compact charges}\label{CCsect}
\begin{figure*}
\begin{minipage}{0.5\textwidth}
\centering
\includegraphics[width=0.75\textwidth]{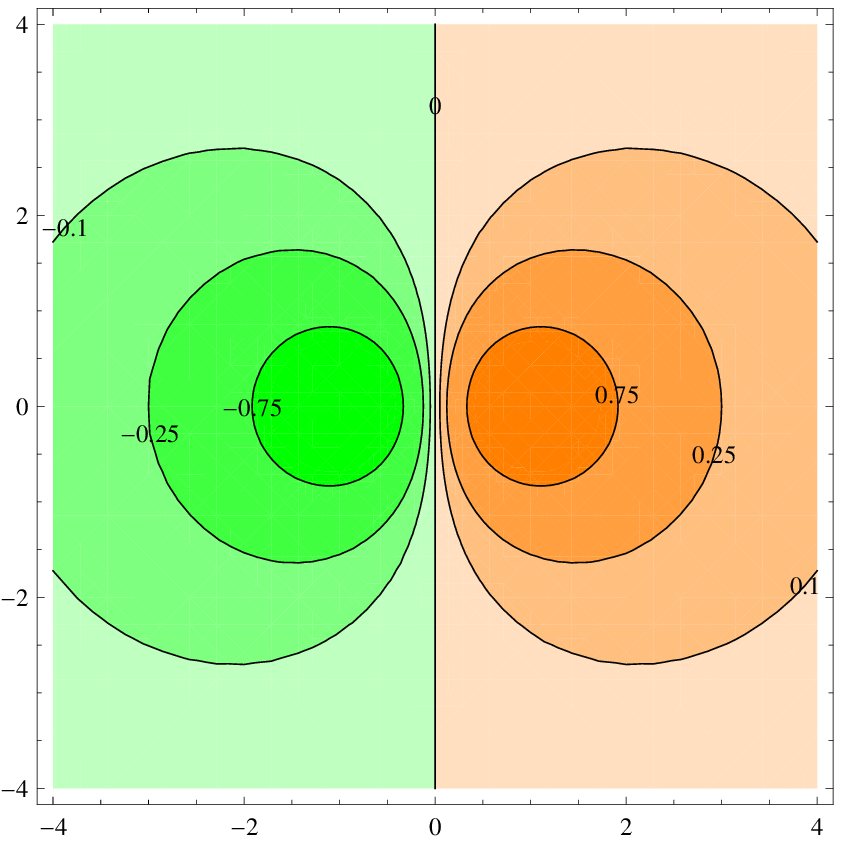}
\caption{Contour plot of the Coulomb scalar potential for two asymptotic charges: green negative, orange positive values.}
\label{asymptpic}
\end{minipage}
\hspace{0.5cm} 
\begin{minipage}{0.5\textwidth}
\centering
\includegraphics[width=0.75\textwidth]{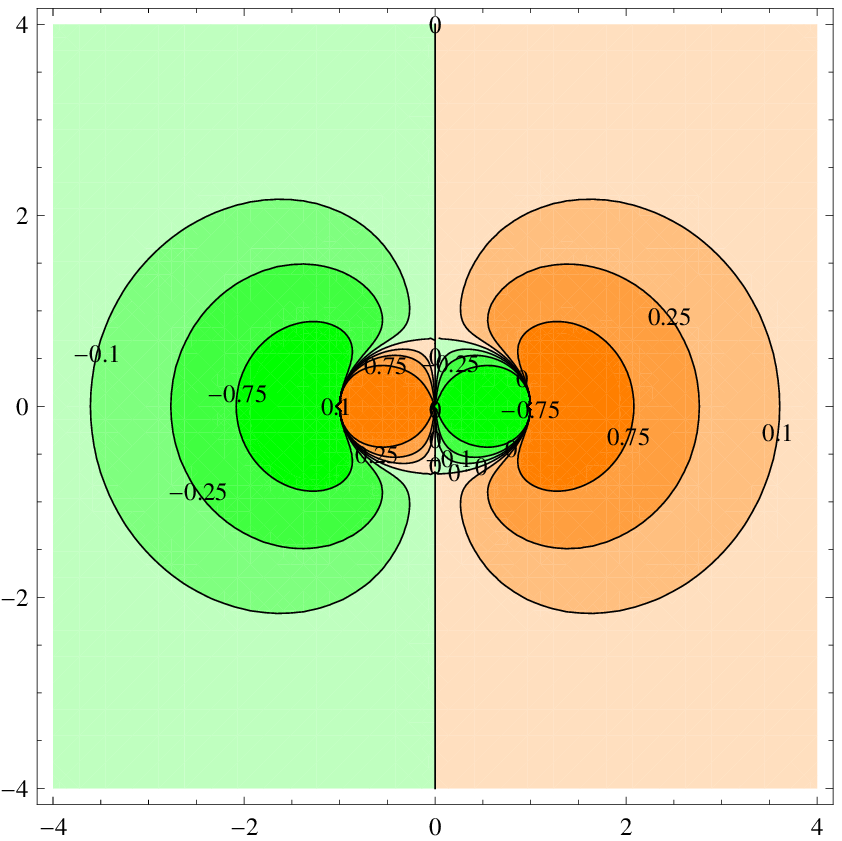}
\caption{Contour plot of the radial electric field for two asymptotic charges: green negative, orange positive values.}
\label{asymptpic1}
\end{minipage}
\end{figure*}
Let us carefully consider the term
\begin{equation}\label{fd}
	\frac{\nabla_jA_j(\a)}{\nabla^2} \equiv \frac{1}{4\pi}\int_{\mathbb{R}^3}\! \ud^3\rv\ \nabla_j\left(\frac1{|\rv-\a|}\right)A_j(\rv)\,,
\end{equation}
which enters into the Coulombic dressed state for a charge at position $\a$. In this section we work in a single time slice and so omit all time dependence. The gauge transformation property of the dressing follows from the replacement of the vector potential $A_j(\rv)$ by $\nabla_j\Lambda(\rv)$ in this expression:
\begin{align*}
  \frac{1}{4\pi}\int_{\mathbb{R}^3}\! \ud^3\rv\ \nabla_j\left(\frac1{|\rv-\a|}\right)A_j(\rv)&\to \frac{1}{4\pi}\int_{\mathbb{R}^3}\! \ud^3\rv\ \nabla_j\left(\frac1{|\rv-\a|}\right)\nabla_j\Lambda(\rv)\\
  &= -\frac{1}{4\pi}\int_{\mathbb{R}^3}\! \ud^3\rv\ \nabla^2\left(\frac1{|\rv-\a|}\right)\Lambda(\rv) = \Lambda(\a)\,,
\end{align*}
The point to note is that there are no surface terms after integrating by parts due to the falloff rate of the fields -- this also means there are no ordering ambiguities between $\nabla_j$ and $1/\nabla^2$ in (\ref{fd}). There is an immediate obstruction, though, to a naive compact version of (\ref{fd}); Gauss's law says a single charge {\it cannot} be compact. However, it offers no such obstruction to compact descriptions of systems which are overall charge--neutral, e.g. those containing an electron--positron pair, and it this which we pursue below.

\begin{figure}[b]
\centering\includegraphics[width=4cm]{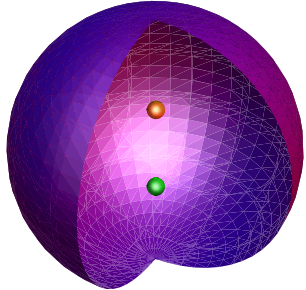}
\caption{Two charged particles within a compact spherical dressing.}
\label{compd}
\end{figure}

To begin, we have from the above that for two (opposite) charges at $\pm\a$, the Coulombic dressing is
\begin{equation}\label{2fd}
    \frac{1}{4\pi}\int_\mathbb{R}\!\ud^3\rv\ \nabla_j\left(\frac1{|\rv-\a|}-\frac1{|\rv+\a|}\right)A_j(\rv)\,,
\end{equation}
resulting in a scalar potential and radial electric field as shown in Fig.'s \ref{asymptpic} and \ref{asymptpic1}.
We now wish to restrict this construction to a finite volume. Gauge invariance alone is not enough to fix the form of the volume, which would in principle be dictated by dynamics. As a natural first step, though, we will simply choose a volume and show how to generalise the Coulombic dressing to fields contained within it. Placing our two charges on the $z$--axis at $\pm\a$, the simplest possible volume maintaining azimuthal symmetry is a sphere, centred on the origin and of radius $R>|\a|$, as depicted in Fig.~\ref{compd}. (In fact, we will later see that a spherical volume appears naturally in models of pair creation and annihilation.) The idea is now to generalise (\ref{2fd}) to a dressing of the form 
\begin{equation}\label{2cd}
    \frac{1}{4\pi}\int_{B}\! \ud V\,\nabla_j\left(\frac1{|\rv-\a|}-\frac1{|\rv+\a|}+f(\rv)\right)A_j(\rv)\,.
\end{equation}
The integration is over the ball $B$ of radius $R$ and $f(\rv)$ is a function to be determined by imposing the required gauge transformation. This function will describe the deviation of the fields from the Coulombic form due to compactness. In order for (\ref{2cd}) to be a dressing we need
\begin{equation}\label{3cd}
    \frac{1}{4\pi}\int_B\! \ud V\,\nabla_j\left(\frac1{|\rv-\a|}-\frac1{|\rv+\a|}+f(\rv)\right)\nabla_j\Lambda(\rv)=\Lambda(\a)-\Lambda(-\a)\,.
\end{equation}
This imposes two conditions. The first is that the function $f(\rv)$ must be harmonic, $\nabla^2 f(\rv)=0$, and the second is that the surface term vanishes when we integrate by parts,
\begin{equation}\label{4cd}
    \int_{S^2}\!\boldsymbol{\ud s}\!\cdot\!\nabla\left(\frac1{|\rv-\a|}-\frac1{|\rv+\a|}+f(\rv)\right)\Lambda(\rv)=0\,.
\end{equation}
As this condition must hold for {\it all} allowed gauge transformations $\Lambda$, we need to solve
\begin{equation}\label{9cd}
    \frac{\partial\ }{\partial r}\left(\frac1{|\rv-\a|}-\frac1{|\rv+\a|}+f(\rv)\right)=0
\end{equation}
on the surface of the ball. This condition says that the radial component, $E_r$, of the electric field produced by the dressing must vanish on the boundary. Recalling that the charge within the ball is the surface integral of the radial electric field,
\begin{equation}\label{10cd}
   \mathcal{Q}=\int_{B}\!\ud V\ \nabla_j E_j=\int_{S^2}\!\ud s\,E_r\;,
\end{equation}
we see that, as a consistency condition, such a compact dressing can {\it only} be constructed if the total charge contained within the dressing region is zero.  (Mathematically, in order to define a compact dressing for $i$ positive and $j$ negative charges at positions $\y_i$ and $\y'_j$ respectively in a general volume $\mathcal V$, we require a Neumann Green's function $G_N$ of the Laplacian which obeys
\be
	\nabla^2_{\z} G_N(\z) = \sum\limits_i \delta^3(\z-\y_i) -\sum\limits_j \delta^3(\z-\y'_j)\quad \in \mathcal{V}\;,\qquad \boldsymbol{\ud s}.\nabla G_N =0 \quad \in \partial \mathcal{V}\;.
\ee
However, integrating $\nabla^2 G_N$ over the volume and using the Neumann condition implies that, for consistency, the total charge contained in $\mathcal V$ must be zero.)

We can now solve for $f(\rv)$. Recall that functions which are both harmonic and azimuthally symmetric have the expansion
\begin{equation}\label{6cd}
    f(\rv)=f(r,\theta)=\sum_{\ell=0}^\infty(\alpha_\ell r^\ell+\beta_\ell r^{-(\ell+1)})P_\ell(\cos\theta)\,,
\end{equation}
in terms of Legendre Polynomials $P_\ell$. As we do not want $f$ to effectively introduce new charges inside the ball, we require the coefficients $\beta_\ell=0$, and therefore
\begin{equation}\label{7cd}
    f(r,\theta)=\sum_{\ell=0}^\infty \alpha_\ell r^\ell P_\ell(\cos\theta)\quad \implies \quad \boldsymbol{\ud s}\cdot\nabla f= \left.\frac{\partial f}{\partial r}\right|_{r=R}=\sum_{\ell=0}^\infty\ell \alpha_\ell R^{\ell-1}P_\ell(\cos\theta)\,.
\end{equation}
Solving (\ref{9cd}) where (\ref{7cd}) holds is straightforward once we recall that, for $a\equiv|\a|<R$,
\begin{equation}\label{11cd}
    \frac1{|\rv-\a|}=\sum_{\ell=0}^\infty \frac{a^\ell}{r^{\ell+1}}P_\ell(\cos\theta)\;,
\end{equation}
and hence that
\begin{equation}\label{12cd}
   \left. \frac{\partial\ }{\partial r}\left(\frac1{|\rv-\a|}-\frac1{|\rv+\a|}\right)\right|_{r=R}=
   -\frac{4}{R^2}\sum_{\ell=0}^\infty(\ell+1)\left(\frac{a}{R}\right)^{2\ell+1}P_{2\ell+1}(\cos\theta)\,.
\end{equation}
From the expansions (\ref{7cd}) and (\ref{12cd}) it is easy to see that our condition (\ref{9cd}) requires $\alpha_{2\ell}=0$ while
\begin{equation}\label{13cd}
    \alpha_{2\ell+1}=\frac4{R^{2\ell+2}}\left(\frac{\ell+1}{2\ell+1}\right)\left(\frac{a}{R}\right)^{2\ell+1}\,,
\end{equation}
and there is no restriction on the constant term $\alpha_0$. The compact dressing (\ref{2cd}) therefore transforms correctly provided
\begin{equation}\label{14cd}
    f(r,\theta)=\frac4R \sum_{\ell=0}^\infty \left(\frac{\ell+1}{2\ell+1}\right) \left(\frac{a}{R}\right)^{2\ell+1} \left(\frac{r}{R}\right)^{2\ell+1} P_{2\ell+1}(\cos\theta)\,.
\end{equation}
To better understand this dressing, we turn now to the fields it generates.
\subsection{The compact potential and fields}
\begin{figure*}
\begin{minipage}{0.5\textwidth} 
\centering
\includegraphics[width=0.75\textwidth]{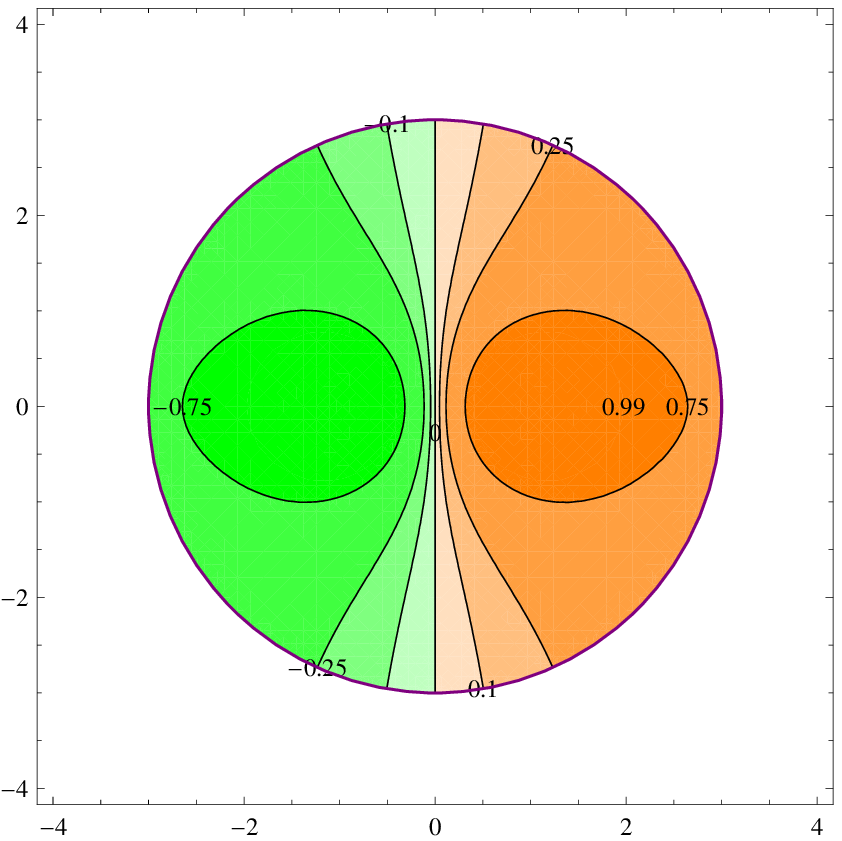}
\caption{Contour plot of the scalar potential for two compact charges at $|\a|=1$ with boundary $R=3$. Note the distortion compared to Fig.~\ref{asymptpic}.}
\label{boxedV}
\end{minipage}
\hspace{0.5cm} 
\begin{minipage}{0.5\textwidth}
\centering
\includegraphics[width=0.75\textwidth]{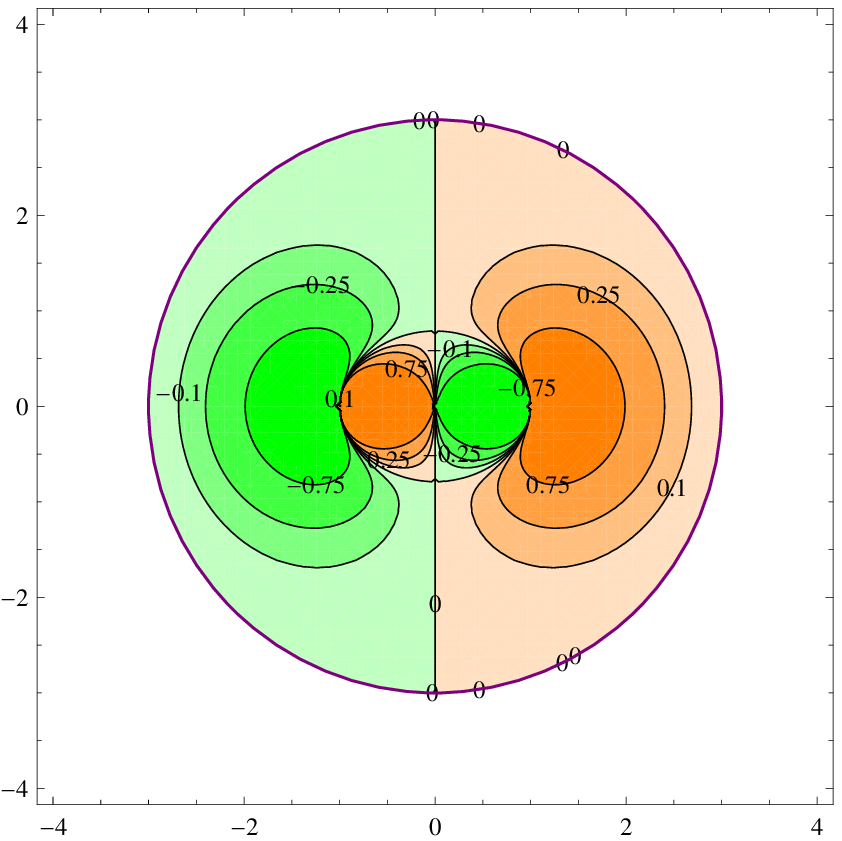}
\caption{Contour plot of the radial electric field for two compact charges at $|\a|=1$, boundary $R=3$. Compare with the asymptotic fields of Fig.~\ref{asymptpic1}.}
\label{boxedE}
\end{minipage}
\end{figure*}
\begin{figure}[ht]
\centering\includegraphics[width=0.4\textwidth]{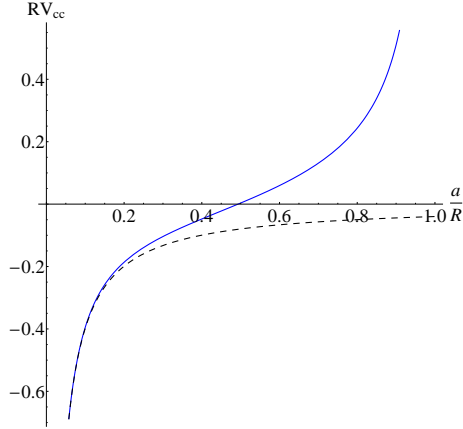}
\caption{The inter--charge potential $V_\text{cc}$ for two compact charges (blue, solid line), as a function of $a/R$. The potential is approximately Coulombic (black, dashed line) if the sources are deep inside the sphere, and diverges if we push the sources to the sphere's boundary.}
\label{Vpot}
\end{figure}

The scalar potential and radial electric fields associated to our compact charges are plotted in Fig.~\ref{boxedV} and Fig.~\ref{boxedE}, respectively, truncated to the interior of the ball. Comparing these plots with Fig.~\ref{asymptpic} and Fig.~\ref{asymptpic1}, the distortion of the scalar potential and field due to the compact form of the dressing is clear to see. In particular, the zero value of the field strength on the boundary of the sphere is clearly pronounced. We can now calculate the potential energy $V_\text{cc}$ between the two charges at positions $\pm\a$. This will confirm two things. First, that the inter--charge potential is greater than the Coulomb potential, signifying an excited state, and second that it is finite, unlike that between charges defined using Wilson lines. Taking the expectation value of the Hamiltonian and dropping self energies, we find the potential between the two charges is
\be
\begin{split}\label{bob}
	V_\text{cc} &= -\frac{e^2}{4\pi}\frac{1}{|\a+\a|} + \frac{e^2}{8\pi} \big(f(\a)-f(-\a)\big) \\
		&=-\frac{e^2}{4\pi}\frac{1}{|\a+\a|} + \frac{e^2}{2\pi} \bigg(\frac{a^2R}{R^4-a^4}+\frac{1}{R}\tanh{}^{-1}\frac{a^2}{R^2}\bigg)\;.
\end{split}
\ee
The second line of the above is calculated by explicitly performing the sums in $f(\a)$. Recalling $a<R$, we see that the second and third term of the above are positive, and so the potential $V_\text{cc}$ is {\it greater} than the Coulomb potential (which appears as the first term of (\ref{bob})), confirming that our compact charges describe an excited state. Importantly, it is {\it finitely} excited, since the potential is manifestly finite. It is plotted in Fig.~\ref{Vpot}. Note that as $R\to\infty$ we smoothly recover the Coulomb potential, and so if $a\ll R$ the potential is approximately Coulombic. If we try to reduce $R\to |\a|$, such that the charges sit on the boundary of the field envelope, a divergence develops in the potential. These statements reflect our earlier observation that it is the Coulomb state, with its infinitely wide dressing, which has minimal energy.
Our construction can be considered as a snapshot in time of some two--charge process. If we were to embed this state into a static charge system, and allow time to evolve, the charges would remain fixed at $\pm\a$ but the bounding sphere would swell out to infinity, and the harmonic function $f(\rv)$ would be suppressed, eventually leaving only the asymptotic Coulomb dressing.

One of our motivations for the above discussion was that, in pair creation, the fields of the newly formed charges are constrained by causality; our ability to confine the fields of our charges to a finite volume bodes well for an application of the dressing approach to this process. Before we can study this in more detail, though, we need to understand the dressings for moving charges. In the following section we go to a Hamiltonian picture, and use it to demonstrate explicitly that, firstly, the ground state of a static charge is given by the Coulombic dressing and then, in the following section, we generalise this calculation to moving charges. We will then be in a position to model pair creation and annihilation.
\section{Physical charges from a variational principle}\label{VPsect}
In the above, we singled out the Coulomb dressing for special attention, and noted that it gave static charges of minimal energy. We make this statement explicit below, and give the necessary groundwork for generalising the Coulombic dressing to moving charges, which we will address in the next section.

We continue to work with (infinitely) heavy charges in order to keep the presentation simple and explicit. This can either be understood as a toy model for QED, or as essentially restricting full QED to the asymptotic regime in which the electron mass gives the largest scale in the game (and all particles are well separated \cite{Horan:1999ba}). We will work with either one or two heavy charges of distinct and definite four--velocities $v^\mu$ and $u^\mu$, the action of which is given by the leading order `HQET' action \cite{Georgi:1990um, Grozin:2004yc}
\be\label{lag}
	\mathcal{L}=-\frac{1}{4}F^{\mu\nu}F_{\mu\nu} + i Q^\dagger_v v^\mu (\partial_\mu + ie A_\mu)\,Q_v +Q^\dagger_u u^\mu (\partial_\mu - ie A_\mu)\,Q_u\;.
\ee
Here $Q_v$ and $Q_u$ are heavy matter fields and we have assigned them opposite but equal charges. (The HQET action should include an integral over velocities, but these are superselected and since we will have charges of distinct velocities, we can drop the integral.) We take $v^\mu=\gamma_v(1,\vv)$, $u^\mu=\gamma_u(1,\uu)$, but we will drop all the velocity subscripts when only one charge is present. 

We will work in a Hamiltonian picture, the application of which to Yang--Mills theory has seen a recent renewal of interest \cite{Reinhardt:2008ax, Reinhardt:2008ij, Campagnari:2009km}. The Hamiltonian is (in terms of vectors $A^j$, $\mathrm{v}^j$, $\mathrm{u}^j$ and derivatives $\nabla_j$)
\be\label{H}\begin{split}
	H = \int\!\ud^3x\ \frac{1}{2} \underbrace{(\E^2 + \B^2)}_\text{Free photons} -\underbrace{e\gamma_v\, \vv\!\cdot\!\A \,Q_v^\dagger\,Q_v+e\gamma_u\, \uu\!\cdot\!\A \,Q_u^\dagger\,Q_u}_\text{Interactions} +\underbrace{i\gamma_v\,(\vv\!\cdot\!\nabla Q^\dagger_v)\, Q_v +i\gamma_u\,(\uu\!\cdot\!\nabla Q^\dagger_u)\, Q_u}_\text{Free matter}\;.
\end{split}\ee
We take our matter fields to be bosonic for simplicity of presentation, but our method and conclusions are equally applicable to fermionic fields. Indeed, the interaction Hamiltonian in the asymptotic regime of scalar QED is essentially the same as that of ordinary, fermionic QED, and in particular the infra--red structure of the two theories is the same. Note that for static matter, (\ref{H}) becomes the free photon Hamiltonian.

The U(1) + heavy matter theory, defined by (\ref{lag}) or (\ref{H}), can be solved exactly. Its nontriviality is due, in effect, to $A_0$. In a Hamiltonian picture, $A_0$ appears only as a Lagrange multiplier, and implies that along with the Schr\"odinger equation, states must also obey Gauss's law,
\be\label{gauss}
	\nabla\!\cdot\!\E+ e\gamma_v \, Q_v^\dagger\, Q_v -e\gamma_u \, Q^\dagger_u\, Q_u=0\;,
\ee
which commutes with the Hamiltonian and is the condition that the state be gauge invariant\footnote{We are not fixing a gauge here. As a result, expectation values of observables contain a divergence. However, this always factors out into a volume (of the gauge group), which can be absorbed into state normalisation and therefore discarded.}. 
The physical content of the theory is given by first writing down all states which obey Gauss's law, and then imposing the Schr\"odinger equation on these. This gives us all admissible states, and contains enough freedom to, for example, set up the string--like system described above in an initial time slice, and then to let it evolve in time. However, we are interested in finite energy states, in particular the ground state. To identify it, we take the general state $\Psi$ obeying both Gauss's law and the Schr\"odinger equation, and minimise the expectation value $\langle H \rangle_\Psi$. (For static charges it is perhaps more natural to solve the time independent Schr\"odinger equation. However, when we consider moving charges we will be forced to consider the time {\it dependent} Schr\"odinger equation, in which case the minimisation approach is more applicable.)

We work for the remainder of the paper in a Schr\"odinger representation in which we diagonalise $A^j$ and $Q^\dagger$ on the quantisation surface and take
\be
	\E(\z) = -i \frac{\delta}{\delta \A(\z)}\;,\qquad Q(\z) =  \frac{1}{\gamma}\frac{\delta}{\delta Q^\dagger(\z)}\;.
\ee
States will be represented by wavefunctionals of $\A$ and $Q^\dagger$. In all our states, charges sit in the photon vacuum $\Psi_0$,
\be\label{vac}
	\Psi_0[\A;t] = \exp\bigg[-i E_0 t - \frac{1}{2} \int\!\ud^3 \z\ \A_T(\z)\!\cdot\!\sqrt{-\nabla^2}\,  \A_T(\z)\bigg]\;.
\ee
Note though, that the operators which excite charged particles from the vacuum are not $\hat{Q}$ or $\hat{Q}^\dagger$ alone, as these are not gauge invariant. It is easy to see that attempting to describe a charged state at position $\y$ by taking the product $Q^\dagger(\y)\Psi_0[\A;t]$, the resulting state is not invariant under the gauge transformations
\be\label{trans}
	Q^\dagger(\z) \to \e^{ie \Lambda(\z)}Q^\dagger(\z)\;,\qquad \A(\z)\to \A(\z)-\nabla \Lambda(\z)\;.
\ee
This is, of course, where Gauss's law is required, and this will generate our dressing. We now show this explicitly for the case of a static charge.
\subsection{Static charges}
For static charges, the Hamiltonian (\ref{H}) reduces to that of free photons, but we still have Gauss's law to impose. Considering a state $\ket\Psi$ with matter content $Q^\dagger(\y)$, i.e. a single static source at position $\y$, Gauss's law acting on the state wavefunction becomes
\be\label{inc1}
	-i\nabla_j\frac{\delta}{\delta A^j(\z)}\Psi+ e\delta^3(\z-\y)\Psi=0 \quad \implies \quad \Psi = Q^\dagger(\y)\exp\bigg[ ie\frac{\nabla\cdot\A(\y)}{\nabla^2}\bigg]\times \tilde\Psi[\A_T;t]\;,
\ee
where $\tilde\Psi$ is an arbitrary function of $\A_T$, the transverse, gauge invariant photon field. We see that the nonlocal Coulombic term has emerged immediately as a natural consequence of Gauss's law. In fact this determines the longitudinal content of the state completely. As before, the interpretation of this addition is that a cloud of longitudinal gauge bosons is required in order to compensate for the transformation of the matter field. Note that only the cloud and matter {\it together} is locally gauge invariant under (\ref{trans}). 

Demanding additionally that the Schr\"odinger equation is satisfied requires adding the photon vacuum, and allows us to add a further transverse component to the state. The general solution is\footnote{In order to keep the discussion as concise as possible, we restrict to the case where $\tilde\Psi$ is a general Gaussian, which is appropriate here since we are only interested in ground state wavefunctionals. All our solutions are still exact.}
\be\label{cod}
	\Psi = Q^\dagger(\y) \exp\bigg[ -i\epsilon(t) + ie \int\!\ud^3\z\ \chi(\z,t)\cdot \A(\z)\bigg]\Psi_0[\A_T;t]\;.
\ee
Here $\chi$ is the dressing, representing the photon cloud, and contains two pieces; that part which is fixed by Gauss' law, (\ref{inc1}), and an additional transverse part allowed by the Schr\"odinger equation. In co--ordinate and momentum space the dressing is
\be	
	\chi^j(\z,t) = -\frac{1}{4\pi}\frac{z^j-y^j}{|\z-\y|} + \omega^j_0(\z,t) \equiv  \int\!\frac{\ud^3p}{(2\pi)^3}\ \frac{ip^j}{|\p^2|}\,e^{i\p.\z-\p.\y} + \tilde{\omega}_0^j(\p)\,e^{i\p.\z-i|\p|t}
\ee
where $\omega_0$ is defined as shown by the arbitrary {\it transverse} function $\tilde\omega_0(\p)$. Also, $\dot\epsilon(t)=\tfrac{e^2}{2}\int\!\ud^3\z\ \chi^2$, which is required to solve the Schr\"odinger equation.

Our problem is now to identify the correct $\tilde\omega_0$ which gives us a physical asymptotic charge. Gauge invariance is already assured, but this is not enough to fix the physics of the state which should, for example, reproduce the known electromagnetic fields of a static (or, later, moving) particle. We could try to take this as a criterion for identifying the correct state, but this is not easily extended to other situations where we do not have previous intuition to draw on -- and of course, showing that such fields arise naturally is one of our goals. We must appeal to another principle, and the only natural, and immediately generalisable, candidate is that the physical state should be that of minimal energy, i.e. the state to which the system will relax at asymptotic times.  Therefore, we will minimise the expectation value $\langle H \rangle_\Psi$ in the state (\ref{cod}).  As the wavefunction is Gaussian,  calculating the expectation value of $H$ in the state $\Psi$ is a simple matter. We find
\be\label{sola}
	\langle H \rangle_\Psi = \frac{e^2}{2}\int\!\frac{\ud^3p}{(2\pi)^3}\ |\tilde\omega_0(\p)|^2 + \text{constant}\;.
\ee
We see that the presence of any transverse term in the dressing only {\it raises} the energy -- the ground state (which is an eigenstate here as there is no explicit time dependence) is therefore given by taking $\tilde\omega_0=0$ in (\ref{cod}), which leaves us with a Coulombic charge. We have really imposed nothing but gauge invariance to obtain this solution, and we conclude that the stable, minimal energy states of charged particles are described by nonlocal Coulombic dressings. Although there are screening corrections to this calculation, they are generated by light flavours, and so will not be addressed here -- see \cite{Bagan:2001wj}. The above calculation is easily extended to states of multiple charges. If two charges are present, for example, the ground state energy -- the `constant' in (\ref{sola}) -- contains the inter--fermion Coulomb potential (\ref{VC}), and of course taking the expectation values of the electric field $\langle \E\rangle_\Psi$ generates the electric field (\ref{El}), plotted in Fig.~\ref{asymptpic1}.

In the next section we generalise the minimal energy dressing to moving charges (see also \cite{Bagan:1999jf}), but before proceeding, we make a brief mention of the normalisation of states in the matter sector. The inner product between two states, $\ket\Phi$ and $\ket\Psi$, in our antiholomorphic representation of the matter fields \cite[Chapter 9-1-2]{Itzykson:1980rh}, contains 
\be\label{IP}
	\int\!\pathD Q\pathD Q^\dagger\ \Phi^\dagger[Q] \ \e^{-\gamma\int\!\ud^3\x\, |Q(\x)|^2}\ \Psi[Q^\dagger] \;,
\ee
which in the above calculation (with $\gamma=1$ for static charges) generates a $\delta^3(0)$ in both the norm of the state $\Psi$ and in $\langle H \rangle_\Psi$. This is really due to the sharp velocities of the heavy matter, and the divergence can be included in the normalisation of the state and so discarded. It is important to note that such terms are present, as dealing with them when our charges are moving is a little more subtle, as we will see.
\section{Moving charges}\label{Msect}
In this section we consider moving charges, and construct their ground state, in analogy to the discussion of static charges above. We will again focus on a single charge, and we will again see that it is described by dressed matter. We will show explicitly that this asymptotic dressing gives the correct electromagnetic fields of a moving charge. Note that for a single charge we could boost to a frame in which $v^\mu=(1,\boldsymbol{0})$, and in this frame the matter fields are equivalent to static sources. However, in more general situations with multiple moving charges, there will not exist a frame in which all 3--velocities vanish, so it is important to understand the $\vv\not=0$ case. We begin with a very brief discussion of the free heavy matter theory, which is necessary now that the matter fields are dynamical.
\subsection{The free matter theory}
As heavy particles only propagate forward in time, there are no loops and therefore no sea of virtual particles making up a ``heavy charge vacuum", which is literally empty space. Instead, states $\qket\Psi$ obeying the free matter Schr\"odinger equation\footnote{We use $\qket{\cdot}$ to denote states containing {\it only} free heavy matter.},
\be
	i\frac{\partial}{\partial t}\,\qket\Psi = \int\!\ud^3x\ i\gamma(\vv.\nabla Q^\dagger)Q\, \qket\Psi
\ee
(see the final terms of (\ref{H})) describe particles traversing paths parallel to the line $\vv t$. The state $\qket{{\y_t}}$ which describes a single particle originating at position $\y$ and traversing the path ${\y_t}:=\y+\vv t$ has wavefunctional
\be\label{1}
	\qbracket{Q^\dagger}{{\y_t}} \equiv \int\!\ud^3\x\ \delta^3(\x-{\y_t})\ Q^\dagger(\x) = Q^\dagger({\y_t})\;.
\ee
From (\ref{IP}) we can calculate the norm of the states (\ref{1}), and expectation values of the Hamiltonian, which will be important later on,
\be
	\qbracket{{\y'_t}}{{\y_t}} = \frac{1}{\gamma}\delta^3({\y'_t}-{\y_t})=\frac{1}{\gamma}\delta^3({\y'}-\y)\;,\qquad (\,{\y'_t}\,|H|\,{\y_t}\,) = \frac{i}{\gamma}\vv.\nabla_{\y}\delta^3({\y'}-\y)\;.
\ee
\subsection{The ground state of moving charges}
Turning the gauge fields back on, our heavy charges now sit in the photon vacuum $\Psi_0$ introduced in (\ref{vac}).  For a state with the matter content of (\ref{1}), the exact solution to both Gauss's law and the Schr\"odinger equation is
\be\label{sol}
	\Psi[\A,Q^\dagger; t] = Q^\dagger(\y_t)\exp\bigg[ -i\epsilon(t) + ie \int\!\ud^3z\  \chi(\z,t) \!\cdot\!\A(\z)\bigg]\Psi_0[\A;t]\;,
\ee
in complete analogy to (\ref{cod}). In this system the dressing is
\be\label{chiv}
	\chi^j(\z,t) = \int\!\frac{\ud^3 p}{(2\pi)^3}\ \tilde{\chi}^j_v\,\e^{i\p.\z-i\p.\y_t} + \tilde\omega^j\,\e^{i\p.\z-i|\p|t}\;,
\ee
where $\chi_v$ is that part which is fixed -- it contains the necessary Coulombic piece for a particle at position $\y_t$, and now an additional transverse contribution fixed by the Schr\"odinger equation,
\be
	\tilde{\chi}_v \equiv\frac{i\p}{|\p|^2}-\frac{i\vv_T}{|\p|-\vv.\p} = \frac{i}{|\p|-\vv.\p}\bigg( \frac{\p}{|\p|}-\vv\bigg) \;.
\ee
The remaining term of the dressing, $\tilde\omega(\p)$, is transverse, but again is otherwise arbitrary. It is easily checked that the state $\Psi$ is invariant under the gauge transformations (\ref{trans}). We now minimise the Hamiltonian. We will write $\ket{\Psi;t}$ for the state corresponding to (\ref{sol}). This can be decomposed into a product $\qket{\y_t}\ket{\y_t}$, where the first ket is the free matter state, as before, and the second is the photon state. Acting on $\ket{\Psi;t}$, we can then decompose the Hamiltonian into $\hat H\equiv \hat{H}_A + \hat{H}_Q$, where $\hat{H}_Q$ is the free matter Hamiltonian and $\hat{H}_A$ contains the remaining terms in (\ref{H}), using Gauss's law to evaluate $Q^\dagger Q$, i.e.
\be
	\hat{H}_A = \int\!\ud^3x\ \frac{1}{2} {\E}^2+ \frac{1}{2}\B^2+e\, \vv\!\cdot\!\A \,\delta^3(\x-\y_t)\;.
\ee
As before, calculating the expectation value is straightforward since our exact wavefunctional is Gaussian, but for one subtlety not encountered for static charges. The normalisation of the states (\ref{sol}) contains a $\delta^3(\y_t-\y_t)=\delta^3(0)$ coming from the matter sector, see (\ref{IP}), but when acting with the Hamiltonian we now also pick up terms with {\it derivatives} $\vv\!\cdot\!\nabla\delta^3(0)$. In order to make sense of sums of such terms, we must displace the charge in the dual state\footnote{Without this, results are nonsensical -- for example, the expectation value of $H$ fails to be  manifestly time independent.} $\bra{\Psi;t}$ from the path $\y_t$ to the path $\y'_t \equiv \y'+\vv t$, for some $\y'$. With this regulator in place, we calculate
\be\label{do1}
	\bra{\Psi';t}\hat{H}\ket{\Psi;t} = \delta^3(\y'-\y)\,\bra{\y'_t}\hat{H}_A\ket{\y_t}+    i\,\bracket{\y'_t}{\y_t}\, \vv.\nabla \delta^3(\y'-\y)\;.
\ee
The regulator $\y'$ is needed to make sense of the second term of this expression, in which the derivative $\vv.\nabla$ may be taken with respect to $\y$. In order to separate the physically interesting part of the expectation value from these delta functions, we replace (\ref{do1}) with its equivalent, in the distribution sense, form\footnote{The same result would be obtained by considering position--space wave packets of charges, where the integral appearing in the wave packet would allow us to integrate by parts in (\ref{do1}).}
\be\label{do2}
	\bra{\Psi;t}\hat{H}\ket{\Psi;t}  = \delta^3(\y'-\y) \bigg[ \bra{\y'_t} \hat{H}_A \ket{\y_t} -  i\, \vv.\nabla\, \bracket{\y'_t}{\y_t} \bigg]\bigg|_{\y'=\y}\;.
\ee
We see that extremising the Hamiltonian is equivalent to extremising the term in square brackets above, as everything else is fixed, i.e. we really extremise an effective Hamiltonian which sees only the photonic sector of the theory. Now that we understand the delta functions, they may be treated as a normalisation, as before, and we find that the expectation value is, in analogy to (\ref{sola}),
\be\label{en}
	\bra{\y_t}\big[\hat{H}_A - i\, \vv.\nabla_{\y}\big]\, \ket{\y_t}= \frac{e^2}{2}\int\!\frac{\ud^3p}{(2\pi)^3} \ |\,\tilde\omega(\p)|^2\ +\ \text{constant.}
\ee
It is clear that our effective Hamiltonian is bounded below, and in particular we see that contributions from the undetermined part of the dressing, $\tilde\omega$, can only {\it raise} the energy of the state. The conclusion is therefore that the minimal energy state has $\tilde\omega=0$, and is described by matter surrounded by the boson cloud $\chi_v$ introduced in (\ref{chiv}), which contains both longitudinal components and transverse components going like $\vv_T$. What is the physics of this state? The electromagnetic fields it generates are given by taking expectation values of the operators $\E$ and $\B$. We find
\be\begin{split}\label{fields}
	\langle\E(\x,t)\rangle_\Psi &= -\frac{e\gamma}{4\pi}\frac{\x-\y_t}{\big[|\x-\y_t|^2+\gamma^2(\vv.(\x-\y_t))^2\big]^{3/2}}\;, \\
	\langle\B(\x,t) \rangle_\Psi &= \vv\times \langle\E(\x,t)\rangle_\Psi\;.
\end{split}\ee
These are exactly the fields of a particle of charge $e$ and velocity $v^\mu$ \cite{Jackson}.  As for the static case, we have found that the physical description of moving charges is of a gauge invariant, nonlocal composite of matter and a boson cloud. Again, despite the nonlocality of the state, the electromagnetic fields it generates are good, local, observables which describe the correct physics. In order to further understand the nature of these physical charges, we return briefly to the free heavy matter theory, from which we will show that the energy we have minimised is that part which goes into changing the form of the fields, beyond what is necessary to move the dressing along with the matter.
\subsection{Covariantly constant fields}
The equations of motion of the free matter field are, from (\ref{lag}),
\be\label{cc}
	v^\mu\partial_\mu Q(y) \equiv \gamma\bigg(\frac{\partial}{\partial y^0} + \mathrm{v}^j\frac{\partial}{\partial y^j}\bigg)\,Q(y)=0\;,
\ee
which is the statement that the field is covariantly constant along worldlines with tangent vector $v^\mu$ (the same applies to the conjugate field $Q^\dagger$). If we parameterise using $t\equiv y^0$, these worldlines are of course the paths $\y_t=\y+\vv t$ considered previously. Naturally, the quantum Heisenberg operators $\hat{Q}(t,\y)$ and $\hat{Q}^\dagger(t,\y)$ also obey (\ref{cc}). It is worth understanding how the condition (\ref{cc}) appears in the Hamiltonian picture, as it will reveal something about our physical charge solutions. From (\ref{1}) we could write the Schr\"odinger equation for the one--particle state as
\be\label{0trans}
	\frac{\partial}{\partial t}\ \qbracket{Q^\dagger}{\y_t} = +\mathrm{v}^j\frac{\partial}{\partial y^j}\qbracket{Q^\dagger}{\y_t}\;,
\ee
which makes it clear that time translation in the free theory is just translation along heavy particle worldlines, but (\ref{0trans}) cannot immediately be associated with the condition of constant covariance because it has the {\it wrong} sign.  Rather, solving the Schr\"odinger equation requires that the {\it kernels} appearing in the state are covariantly constant. We can see this quite generally: consider the arbitrary matter wavefunctional
\be
	\int\!\ud^3x_1\ldots \ud^3 x_n\ \Gamma(x_1,\ldots, x_n;t)\ Q^\dagger(x_1)\ldots Q^\dagger(x_n)\;,
\ee
imposing the Schr\"odinger equation on which is equivalent to the condition
\be\label{GenCC}
	\frac{\partial}{\partial t} \Gamma(x_1,\ldots x_n;t)  = -\sum\limits_{k=1}^n\bigg[\mathrm{v}^j \frac{\partial}{\partial x^j_k}\bigg]\Gamma(x_1,\ldots x_n;t)\;.
\ee
Note the leading minus sign on the right hand side. For $n=1$ this is just the previous condition of constant covariance, and it is clear to see from (\ref{1}) that this is obeyed by the kernel in the state $\qbracket{Q^\dagger}{\y_t}$,
\be
	\bigg(\frac{\partial}{\partial t} +\mathrm{v}^j\frac{\partial}{\partial x^j}\bigg)\delta^3(\x-\y_t)=0\;.
\ee
For $n>1$, (\ref{GenCC}) is a generalised covariance condition, and is equivalent to demanding that all heavy charges move along the correct worldlines. We can now see another reason for $\tilde\omega=0$ being the physically stable solution in the interacting theory; it is the only solution to the Schr\"odinger equation which obeys the condition of constant covariance (\ref{GenCC}) on all the kernels in the state (for both the gauge bosons and the matter). Thus, it is the only dressing which preserves the free equations of motion in the interacting theory -- this is the condition which was originally used to construct moving dressings \cite{Bagan:1999jf}. Since the fields (\ref{fields}) are covariantly constant,  setting $\tilde\omega=0$ is also the only way to produce the correct electromagnetic fields. This shows that the calculation in (\ref{en}) should then be viewed as minimising that part of the energy which does {\it not} go into simply translating the state along the heavy particle worldline; i.e. we minimised that part of the energy which goes into changing the {\it form} of the electromagnetic fields.
\subsection{The fields of pair creation}
We now apply our moving dressings to a model of pair creation, considering  a pair of heavy particles, of velocities $\pm\vv$, `created' at the origin $\y=0$ at time $t=0$. As the charges initially sit at the same position, gauge invariance is satisfied and there will be no longitudinal fields around the pair in this time slice, the instant of their `creation'. Extending the construction of the previous sections to states of multiple charges is straightforward. We now use the available degree of freedom $\tilde\omega$, which appears after solving the Schr\"odinger equation, to impose the natural boundary condition that there are also no transverse fields, and hence no fields at all, when $t=0$. The dressing for the two charges at all times $t\geq0$ is then
\be\label{pairdress}
	\chi^j_{pc}(\z,t) = \int\!\frac{\ud^3p}{(2\pi)^3}\,\e^{i\p.\z}\,\bigg[ \tilde{\chi}_{v}^j\ \e^{-i\vv.\p\, t}
 - \tilde{\chi}_{-v}^j\ \e^{i\vv.\p\, t} + i\mathrm{v}_T^j\ \frac{2|\p|\e^{-i|\p|t}}{|\p|^2-(\vv.\p)^2}\bigg]\;.
\ee
It is easily checked that the dressing vanishes at $t=0$. In order to demonstrate that this dressing gives a state obeying causality, we will calculate the electric field of our pair, assuming for simplicity that they are slowly moving, so $|\vv|\ll 1$, and we will retain terms only of order $\vv$. In (\ref{pairdress}), we can expand the denominators to first order in $\vv$ even under the integral, since $\vv.\p/|\p|$ is always less than one. We cannot expand the exponentials without first performing the Fourier integral -- this is because the exponentials see scales other than $|\vv|$, generated by the spacetime position $(t,|\z|)$ of the probe/ observer. The integral to be performed is
\be\label{E-PC}
	\langle\E(\z,t)\rangle_\text{pc} 
	=2e\int\!\frac{\ud^3p}{(2\pi)^3}\,\e^{i\p.\z}\, \bigg[\vv\ \frac{1}{|\p|}\sin|\p|t +\p\frac{1}{|\p|^2}\ \big(\sin \vv.\p\, t - \frac{\vv.\p}{|\p|}\sin |\p|t\big)\bigg]\;.
\ee
We label the three terms appearing in the integrand by \ding{172}, \ding{173} and \ding{174} respectively. The first is simplest to compute,
\be
	\text{\ding{172}} = \frac{e}{2\pi|\z|}\vv\ \delta^3(t-|\z|)\;.
\ee
This term provides the boundary of the volume around the pair, at time $t$, outside of which the electromagnetic fields are zero. The boundary is a spherical shell moving away from the origin at the speed of light -- this is of course precisely the causal boundary of the region into which information from the `creation' event at the origin can propagate to in time $t$. That the field strength is infinite on the boundary is a consequence of placing the charges initially at the same point -- separating the initial positions of the charges can remove this divergence, as illustrated by the compact dressing discussed for static sources in Sect.~\ref{CCsect}.

We next calculate \ding{173} and \ding{174} by first extracting the momentum vectors as derivatives with respect to $\z$, and then performing the scalar Fourier integrals. This leaves us with
\be\label{both}
	\text{\ding{173}} = -\frac{e}{4\pi}\nabla_{\z} \bigg( \frac{1}{|\z+\vv t|} - \frac{1}{|\z-\vv t|}\bigg)\;,\qquad \text{\ding{174}} = \frac{e}{2\pi}\nabla_{\z}\, \vv.\nabla_{\z}\ \frac{\text{min}(t,|\z|)}{|\z|}\;.
\ee
The aim is to confirm that the fields vanish outside the causally allowed region. Since \ding{172} exists only on the causal boundary, the terms in (\ref{both}) must, at time $t$, cancel exactly for an observer sitting outside the ball of radius $t$. We therefore take $|\z|>t$ and, recalling that we are working to first order in $\vv$, find
\be\label{2app}
	\text{\ding{173}} = \frac{e t}{2\pi}\bigg( \frac{1}{|\z|}\,\vv - 3 \frac{\vv.\z}{|\z|^5}\,\z\bigg) = - \text{\ding{174}}\;.
\ee
This confirms that at time $t$, we have $\langle \E \rangle_\text{pc}=0$ outside the sphere of radius $t$, demonstrating causality of the dressing. The magnetic field may be similarly calculated. We note also that \ding{174} does not contribute to the fields inside the ball, at least for small $\vv$, and that within the ball one can still distinguish different regions with different fields, depending on whether the observer sits between the boundary and the charge, or between the charges themselves. The extension of this calculation to QCD, even in perturbation theory, would be fascinating, as it would allow us to see the distribution and motion of colour within the gluon cloud around moving quarks.
\subsection{The fields of pair annihilation}
Consider now the annihilation of two charges. This is not the same as performing a simple time reversal in the above state. That would describe a pair of charges, initially vastly separated, which, as they come together, automatically contract their fields in around themselves so as to fit into a compact volume, and vanish as the charges meet. In other words, we could start initially with charges which are causally disconnected, but which would {\it know} that they are supposed to collide in the far future. This description is unphysical. Instead, what we imagine is that we have two asymptotic charges which eventually meet and annihilate, say at the origin. After the moment of their annihilation, their longitudinal, Coulombic, fields cancel identically at all points in space, and these fields should remain zero thereafter (recall that the Coulomb field ensures gauge invariance of the charges, which have now vanished). The transverse components of the fields remain at the moment of annihilation, but then these will begin to decay away -- a cavity will form around the origin, expanding outward until eventually all observers agree that there are no fields/ charges remaining.

In fact, we can confirm this picture almost immediately from our previous calculations. We take two {\it asymptotic charges} of velocities $\pm\vv$ and parameterise their worldlines such that they meet at the origin at time $t=0$. The dressing is simply
\be\label{PA1}
	\chi^j_{pa}(\z,t) = \int\!\frac{\ud^3p}{(2\pi)^3}\,\e^{i\p.\z}\,\bigg[ \tilde{\chi}_{v}^j\ \e^{-i\vv.\p\, t}
 - \tilde{\chi}_{-v}^j\ \e^{i\vv.\p\, t}\bigg]\;,\qquad t\leq 0\;.
\ee
At $t=0$, the longitudinal fields vanish everywhere, but the transverse fields remain. The state is of course gauge invariant at $t=0$. If the charges now annihilate, the state will continue to evolve in time, but now under the {\it free photon Hamiltonian} $\tfrac{1}{2}\int\!\ud^3\x\ \E^2+\B^2$. The longitudinal components of the fields {\it do not} reappear under this time evolution. Instead, the ``dressing" $\chi_{pa}$ evolves according to
\be\label{PA2}
	\chi^j_{pa}(\z,t) = \int\!\frac{\ud^3p}{(2\pi)^3}\,\e^{i\p.\z}\,\bigg[ \frac{-2i\vv_T^j|\p|}{|\p|^2-(\vv.\p)^2}\bigg]\,e^{-i|\p|t}\;,\qquad t\geq 0\;,
\ee
which, it may be checked, matches (\ref{PA1}) at $t=0$. (Recall that under a sudden change in the Hamiltonian, wavefunctions are continuous.) Note the similarity to the final term of our pair creation dressing, (\ref{pairdress}). Consider again the electric field at small $\vv$,
\be
	\langle\E(\z,t)\rangle_\text{PA} = - 2e\int\!\frac{\ud^3p}{(2\pi)^3}\,\e^{i\p.\z}\, \bigg[\vv\ \frac{1}{|\p|}\sin|\p|t -\p\frac{1}{|\p|^2}\ \frac{\vv.\p}{|\p|}\sin |\p|t\bigg]\;,\qquad t\geq 0\;,
\ee
and compare with (\ref{E-PC}). There is an overall minus sign difference, and the integrand contains precisely the terms \ding{172} and \ding{174} considered above. We thus see a beautiful symmetry between the processes of creation and annihilation, captured by our dressing. Once again, \ding{172} describes the causal boundary of a spherical shell, expanding outward at the speed of light from the origin. Now, at time $t$, we have already found \ding{174} vanishes {\it inside} the ball of radius $t$ --  we therefore have a cavity expanding outward, inside of which the remaining transverse fields of the charges vanishes. As time evolves, the cavity expands to fill the whole of space, eating up the fields of our annihilated charges as the system decays to its new ground state, which is just the photon vacuum.
\section{Conclusions}\label{Concs}
The conclusion that dressings provide the correct, physical description of charges in QED is inescapable. We have shown explicitly that ground state wavefunctions for both static and moving asymptotic charges are described by dressed states. The physical picture is of a matter particle surrounded by a cloud of `photons', neither of which are individually observable, but which together constitute a gauge invariant, physical particle. This description is nonlocal, which is an immediate consequence of gauge invariance, but observables calculated with our states are manifestly local and correctly reproduce classically expected physics.

We have also discussed excited states of electron--positron systems, in which the fields around pairs of charges are confined to a finite volume. We have used these states to model pair creation and annihilation, and we have seen that our states correctly respect the causality constraints expected of such processes.

In future work we hope to be able to study the time evolution of our compact dressings in more detail, and so disentangle their generic (infrared) structure from the model-dependent (ultraviolet) structure. Everything which we have discussed here may be reproduced in QCD order by order in perturbation theory. In particular, the techniques developed in \cite{Mansfield:1993pd, Mansfield:1995je} could be applied to generalising our Hamiltonian analysis of Sect.~\ref{VPsect}.

We recall that the Gribov ambiguity, by causing a breakdown of single quark dressings at a nonperturbative level, generates confinement \cite{Ilderton:2007qy} (see also \cite{Holdom:2007gx, Reinhardt:2008pr, Holdom:2009ws}). Uncovering the role of Gribov copies in the definition of the moving dressing would be extremely significant for understanding the onset of hadronisation. It may also be possible to access nonperturbative properties of charges by considering dressings at the level of the action. We note that just such a construction was applied to unparticles in \cite{Ilderton:2008ab}
and also seems to be involved in the alternative approach to electroweak symmetry breaking described in \cite{Faddeev:2008qc}.

\subsubsection*{Acknowledgements}
A.~I. thanks Paul Mansfield for very useful discussions, and is supported by IRCSET.

\end{document}